\documentclass[conference]{IEEEtran}
\IEEEoverridecommandlockouts
\usepackage{cite}
\usepackage{amsmath,amssymb,amsfonts}
\usepackage{algorithmic}
\usepackage{graphicx}
\usepackage{textcomp}
\usepackage{xcolor}
\usepackage{subcaption}
\usepackage{url}
\def\BibTeX{{\rm B\kern-.05em{\sc i\kern-.025em b}\kern-.08em
    T\kern-.1667em\lower.7ex\hbox{E}\kern-.125emX}}
\begin{document}

\title{Zydeco-Style Spike Sorting Low Power VLSI Architecture for IoT BCI Implants}

\author{\IEEEauthorblockN{Zag ElSayed$^\ddagger$, Murat Ozer$^\ddagger$, Nelly Elsayed$^\ddagger$, Magdy Bayoumi$^\star$}
\IEEEauthorblockA{\textit{School of Information Technology} \\
	\textit{$^\star$Department of Electrical and Computer Engineering} \\
\textit{$^\ddagger$University of Cincinnati, OH, United States}\\
\textit{$^\star$University of Louisiana at Lafayette, LA, United States}
}
}
\thispagestyle{empty}

\begin{huge}
	IEEE Copyright Notice
\end{huge}

\vspace{5mm} 

\begin{large}
	Copyright (c) 2022 IEEE
\end{large}

\vspace{5mm} 

\begin{large}
	Personal use of this material is permitted. Permission from IEEE must be obtained for all other uses, in any current or future media, including reprinting/republishing this material for advertising or promotional purposes, creating new collective works, for resale or redistribution to servers or lists, or reuse of any copyrighted component of this work in other works.
\end{large}

\vspace{5mm} 

\begin{large}
	\textbf{Accepted to be published in:} IEEE WFIoT-2022; 26 October - 11 November , 2022 - Yokohoma, Japan.
	https://wfiot2022.iot.ieee.org/
	
\end{large}

\vspace{5mm} 


\maketitle

\begin{abstract}
Brain Computer Interface (BCI) has great potential for solving many brain signal analysis limitations, mental disorder resolutions, and restoring missing limb functionality via neural-controlled implants. However, there is no single available, and safe implant for daily life usage exists yet. Most of the proposed implants have several implementation issues, such as infection hazards and heat dissipation, which limits their usability and makes it more challenging to pass regulations and quality control production. The wireless implant does not require a chronic wound in the skull. However, the current complex clustering neuron identification algorithms inside the implant chip consume a lot of power and bandwidth, causing higher heat dissipation issues and draining the implant's battery. The spike sorting is the core unit of an invasive BCI chip, which plays a significant role in power consumption, accuracy, and area. Therefore, in this study, we propose a low-power adaptive simplified VLSI architecture, "Zydeco-Style," for BCI spike sorting that is computationally less complex with higher accuracy that performs up to 93.5\% in the worst-case scenario. The architecture uses a low-power Bluetooth Wireless communication module with external IoT medical ICU devices. The proposed architecture was implemented and simulated in Verilog. In addition, we are proposing an implant conceptual design.

\end{abstract}


\section{Introduction}

Brain Computer Interface (BCI) is a communication system that links a human brain activity to a machine via translating neural electrical activity into computer commands. One of the most important goals of BCI research is to enable the handicapped to control artificial limbs by only thinking about the movement action itself and not the motor actin activation control. BCI can offer near-optimal solutions for controlling prosthetic limbs~\cite{1}. Despite the tremendous research effort in the BCI field, the human usability of BCI implants is still limited to laboratory scenarios due to the usability, power consumption, and mobility limitations.

\begin{figure}
	\centerline{\includegraphics[width=8.7cm, height= 6.5cm]{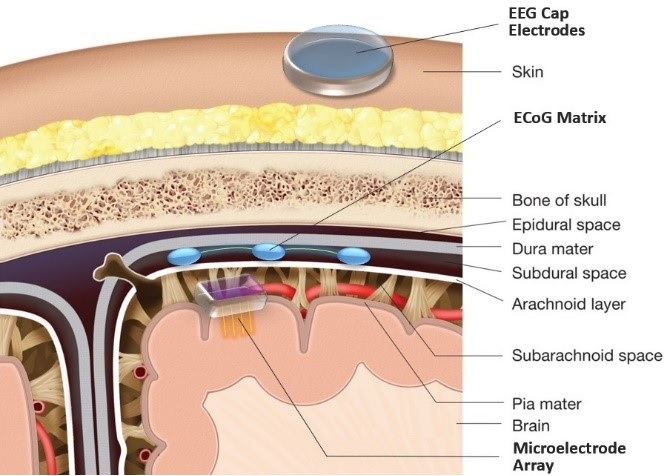}}
	\caption{Cross-section of a top partition of human skull showing the possible positioning of the implant electrodes: EEG caps sensor, ECoG matrix, and Microelectrode array.}
	\label{figure1}
\end{figure}

Electroencephalography (EEG) is a method to record an electrogram of the electrical activity on the scalp that represents the macroscopic activity of the surface layer of the brain underneath. It is typically non-invasive, with the electrodes placed along the scalp. Neural activity sensors are usually classified into three main categories: non-invasive (e.g., EEG), partially invasive (e.g., Electrocorticography), and invasive sensors (e.g., microelectrode array). These sensors can be placed on different layers within the patient's head, such as: on the top of the skull skin, under the skull bone, right over the cortical and the dura tissue, or the Pia Matter itself (gray matter), respectively, as shown in Fig.~\ref{figure1}.

As a matter of fact, there is a trade-off between the electrode sensors' location and the signal-to-noise ratio (SNR). The microelectrode array has the best SNR (at $\mathrm{3.0 \mu Vrms}$)~\cite{3} compared to the EEG cap sensors and the Epidural/Subdural ECoG sensors. On the other hand, it requires the most invasive surgical procedure\cite{4,5,6}. Wireless implants do not need chronic wounds in the subject's skull and provide communication between the BCI Implant (through the low power electromagnetic signaling), where the outside receiver BLE chip could be either located on the prosthetic limb that receives the processed data and executes the movement commands or a medical ICU monitoring device implementing the IoT architecture for medical equipment. It is crucial to notice that while sending each recorded signal of the neuron activity will consume high bandwidth and power, which would conflict with the implant power limitations of the (8-10 mW)\cite{8}. Thus, currently, a signal-processing chip is usually added to the implant as a supporting unit that filters the classified input signal and reduces the amount of data transmitted via a spike sorting module that also lowers the heat dissipation of the implant chip.

Despite that, the BCI implant implementation limitations are challenges~\cite{5}. There are many other issues and challenges that face the common state-of-the-art spike sorting techniques~\cite{5}, especially for VLSI implementation that is discussed in~\cite{7}. Therefore, in this study, we propose a simplified, adaptive, power-efficient BCI spike sorting VLSI architecture, “Zydeco-Style,” for BCI Implants empowered by a low-power BLE technology for wireless communication. The proposed architecture is accurate according to the industrial standards~\cite{5} and follows all the human body implant limitations~\cite{6}.

With the advances in the communication and signal technology of low-power wireless communication, such as the EFR3BG21~\cite{2} allows the medical implants to transmit data and stay connected with a wide variety of medical devices, telemedicine, support IoT architecture, and provide a continue informative status reporting, where the total system draws 5.8mA in TX and 6.8mA in RX while running the Bluetooth (BLE) or near field Communication (NFC) protocol as well as the chip can radiate energy wirelessly to surrounding devices especially, within the ICU monitoring environment using 13.56 MHz field with the help of coils.

The core idea of our proposed system came from the traditional Zydeco music~\cite{7}. Zydeco is a music genre that evolved in southwest Louisiana by French Creole speakers, which blends blues, rhythm and blues, and music indigenous to the Louisiana Creoles and the Native American people of Louisiana. Although it is distinct in origin from the Cajun music of Louisiana, the two forms influenced each other, forming a complex of genres native to the region. The origin of the term "Zydeco" is uncertain. One theory is that it derives from the French phrase: \textit{Les haricots ne sont pas salés}, which, when spoken in Louisiana French, translates as "\textit{the snap beans are not salty}" and is used idiomatically to express hardship.

Zydeco music is typically played in an up-tempo, syncopated manner with a strong rhythmic core. Zydeco music is centered around the accordion, which leads the rest of the band, and a specialized washboard, called a \textit{frottoir}, as a prominent percussive instrument. It blends and fits the solo melody with the base rhythm, producing unique ear-pleasing music. Similarly, in our proposed system, the algorithm divides the input signal immediately at the input channels into two flows of processes that add a second dimension of informative analysis to carry information about the filtered input, as well as the feature classifier for amplitude recognition unit (ARU) and the frequency discrimination unit (FDU) of the stochastic behavior of the neural activity to the spike-sorting algorithm.

\section{Background}
\subsection{Brain Signal Features}

Electroencephalogram was the first recorded in 1924 at Hans Berger's lab in Germany, it was a recording of the electrical activity of the neurons taken from a silver wire penetrated through the patient's skull, and the recorded signal was called alpha waves (8–13 Hz). Later on, other waves were discovered: delta waves (0.5–3 Hz), theta waves (3–8 Hz), beta waves (12–38 Hz), and gamma waves (38–42 Hz). Gamma waves are the fastest of brain waves that are strongly correlated to brain information processing activity~\cite{5}.

\subsection{BCI Signal Processing}

\begin{figure}
	\centerline{\includegraphics[width=8.7cm, height=5.5 cm]{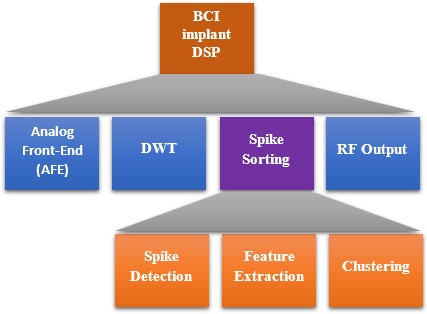}}
	\caption{The block diagram of a BCI implant chip signal processing units, including the spike sorting modules.}
	\label{figure2}
\end{figure}

BCI signal processing unit is usually placed within the implant chip for signal filtering, signal conditioning, neural data identification, and signal parameters dimension reduction. Spike sorting unit performs digital data filtering, feature extracting, and clustering. A block diagram of a common spike sorting BCI module is shown in Fig.~\ref{figure2}.

Spike sorting unit occupies the largest area of the implant chip, and it plays a significant role in power consumption, heat dissipation, and the accuracy of the whole system. Thus, by enhancing the spike sorting module, we can enhance the power consumption reduction. Moreover, by considering reducing the amount of data transmitted, it will benefit the power reduction as well. 

\subsection{Wireless BCI Implant Challenges}

Regarding the physical properties of any BCI implant, it must follow the following limitation~\cite{5}:
\begin{itemize}
	\item Chip area must not exceed 1 $\mathrm{cm^2}$,
	\item Energy dissipation cannot heat the surrounding tissue more than 1 $\mathrm{C^{\circ}}$,
	\item Power consumption must never exceed 10 mW,
	\item The minimum neural identification accuracy must be more than 90\% at 8.4 dB. 
\end{itemize}

In the next section, we will discuss the related proposed BCI architectures for VLSI implementation.

\subsection{Related Work}

Several wireless spike sorting techniques DSP chips were proposed~\cite{10,11}. However, their major issues can be summarized in: battery replacement complexity and implant heat dissipation that causes difficulties for BCI implant safe deployment (e.g., comparing to Cochlear hearing implants~\cite{12}). Several research proposed using Principal Component Analysis (PCA) and Independent Component Analysis (ICA) for an unsupervised, adaptive spike sorting architecture; however, these methods suffer from low accuracy and cache rebuilding downtime. On the other hand, the authors in~\cite{6} proposed a filtered and parameterized K-Means approaches, in addition to proposing a brilliant idea that was implemented on a chip using a backboard-based Spike ID sorter with 16 input channels; however, it requires a prerequisite long training phase.

In the previous work~\cite{13}, we proposed a parameterized artificial immune system for clustering and searching~\cite{8}; where the neural identification task was based on neural fingerprints taken from the dominant channel, and the signal across all the adjacent channels where the system detects (with 100\%) accuracy the neurons at the idea input scenario. Although the system degrades with the increase of the signal-to-noise ratio (SNR), it keeps operating with 91\% accuracy at 15dB; occupies 23\% less area, and decreases bandwidth by an order of magnitude.

\section{Neural Fingerprint}
\begin{figure}
	\centerline{\includegraphics[width=8.5cm, height= 7.5 cm]{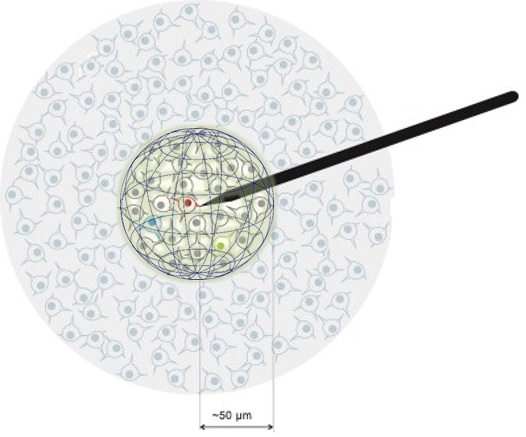}}
	\caption{The diagram of a microelectrode needle tip and the strongest associate noise from the surrounding neurons (imaginary sphere of $r = 50\mu m$).}
	\label{figure3}
\end{figure}

It is important to notice that due to the microelectrode array's physical design and implementation properties, it makes every one second of microelectrode array input recording to contain $\approx$48,000 noise spikes. Moreover, the majority of this strong noise comes from the surrounding neurons to the neuron structure under consideration, which is supposedly connected with the tip of one of the 500 microelectrode needles or, more specifically, to the silver oxide needle, these surrounding neurons can be seen as in a spheric structure around the tip of the needle (an imaginary sphere of $\mathrm{r = 50\mu m})$, shown in Fig.~\ref{figure3}, and that is where most of the attenuation of the neural spike signal is coming from. Additionally, the fact that the amplitude of each of these spikes is usually modeled as coming from a normal distribution with $\mathrm{\mu= 1}$ and $\mathrm{\sigma= 0.2}$.

\begin{figure}
	\centerline{\includegraphics[width=8.7cm, height= 5 cm]{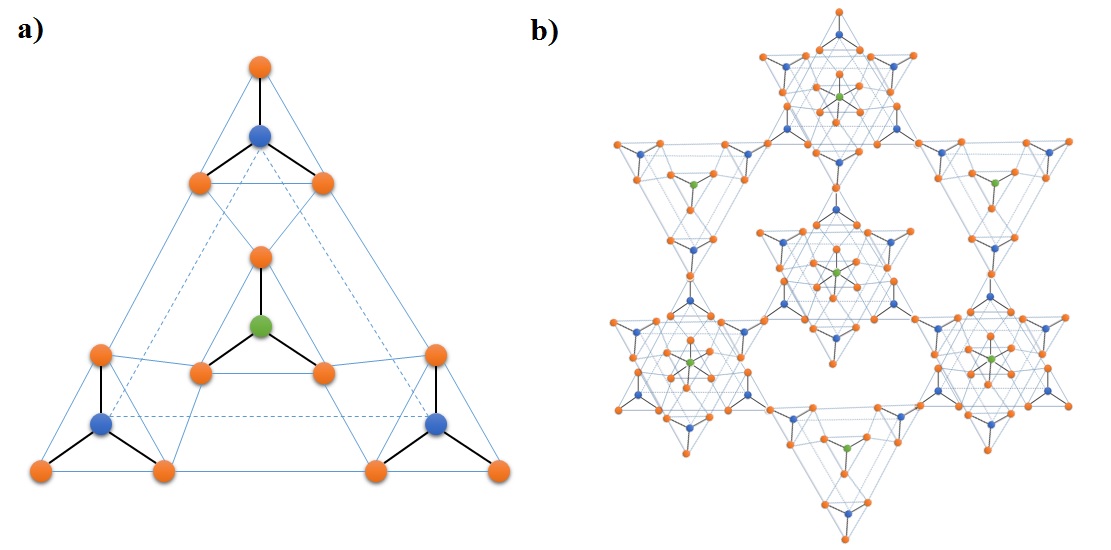}}
	\caption{The novel proposed model of neural noise signaling propagation from the microelectrode needle tip perspective, a) a single neuron and its two types of neighbors as seen from the tip of the needle, b) fractal pattern of the neural circuit shows a connection of four neurons.}
	\label{figure4}
\end{figure}

Thus, we are proposing to use this natural phenomenon of the sensing equipment to describe the firing neuron and produce an individual fingerprint pattern that based on the neural locality principle. Moreover, considering the time delay of the dominant firing neuron and the moment it appears on the closest surrounding channels gives a better vision of the firing neuron. However, to study and analyze this new effect on the neural fingerprint, we had to come up with a new model for neural wiring, following a fractal formation theory, shown in Fig.~\ref{figure4}, a fractal is a never-ending pattern. Fractals are infinitely complex patterns driven by recursion that are self-similar across different scales. They are created by repeating a simple process over and over in an ongoing feedback loop. Fractal patterns are extremely familiar since nature is full of fractals. For instance: trees, rivers, coastlines, mountains, clouds, and hurricanes. 

\begin{figure*}
	\centerline{\includegraphics[width=11cm, height= 6 cm]{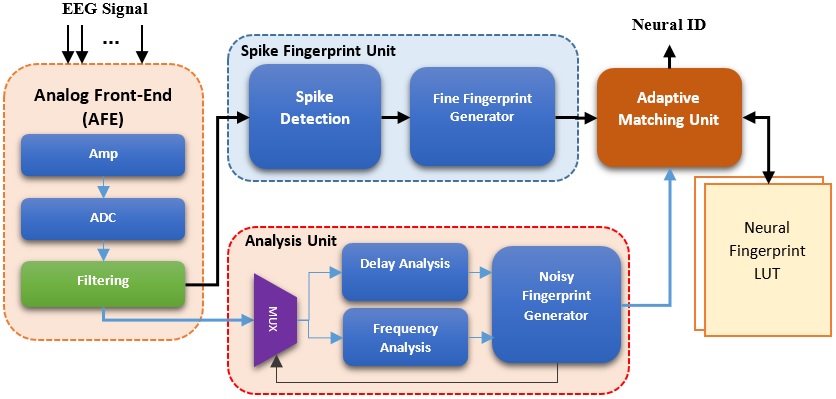}}
	\caption{The block diagram of the proposed Zydeco-Style architecture shows the two main modules for generating neural fingerprints.}
	\label{figure5}
\end{figure*}

The proposed model does not cover the neural fiber connection or the mini-cylinder analysis~\cite{14}. Instead, it describes the generation of the actual signal combined with the surrounding noise propagation path structure of each channel of the microelectrode needles within the sensor coverage space, which is essentially required for the more clarified and sophisticated neural fingerprint creation.

In Fig.~\ref{figure4}, the small circles represent neurons. The blue circle represents the dominant neuron; the surrounding orange circles connected via black edges represent the closest surrounding neurons (the imaginary sphere that usually produces the strongest noise). Therefore, this model does not only offer new circuit connectivity, but it gives a new capability of visualizing the natural neural fingerprint image. Furthermore, this model defines new functionality to the single neuron, regarding the dendrite plasticity due to learning and forgetting because of the long-term excitation (LTE) and long-term potentiation (LTP), respectively, described by the bidirectional model (1) from the Hebbian theory~\cite{13}. 

\begin{equation}
\frac{dW_i(t)}{dt} = \frac{1}{\tau([Ca^{2+}]_i)}(\Omega([Ca^{2+}]_i) -W_i)
\end{equation}
\noindent
where $W_i$ is the weight of the connection for $i$, and $\tau$ is the time constant of the insertion and the removal of the neurotransmitter receptor $Ca^{2+}$, which is the calcium ion concentration, and $\Omega$ is a function of the concentration of calcium ions that linearly depends on the number of receptors on the membrane of the neuron, where the weight propagation still can be calculated by:

\begin{equation}
	\Delta w_{ij} = x_i x_j
\end{equation}
\noindent
where $W_{ij}$ is the weight of the connection for neuron $i$ to neuron $j$ (dendrite connection strength), $x_i$ and $x_j$ are the input for neuron $i$ and neuron $j$, respectively.

Therefore, the proposed model of neural noise signaling/generation behavior is suitable for both: artificial neural network theory and medical brain activity signature specification applications; it provides the functionality and the visual topography of the neural firing and noise distribution operations simultaneously.

\section{Zydeco-Style Architecture}
Based on the facts of the neural activity anatomy, it raises the idea of looking at neural signal recordings from the electrode-based signal acquisition device for the designated neural activity under the study of a specific brain area that is actually blended with the surrounding simultaneous neural activity. Especially if we look at the active potential spiking activity from a wider prospectus, we can see that the attached muscular and vital body-related activities generate simultaneous and asymptotically coherent signals within the same voltage (which in many cases is considered as electroencephalograph noise). However, the actual potential signature of these semi-rhythmic signals can be useful in adding more information details for neural group identification processes, such as neural signature fingerprints. Thus, instead of filtering out this valuable information from the input signal for the spike sorting mechanism (as it is considered by default as useless data in most of the current spike sorting techniques), our proposed method used this information to provide a better classification task. The main principle of the Zydeco Style architecture is to divide the input signal (right after the Analog Front-End (AFE) unit) into two paths, as shown in Fig.~\ref{figure5}. The first path executes the neural identification process via the fine fingerprint generation (FFG) unit implemented in~\cite{13}. The second path connects to the frequency and time analysis process, followed by the global-local fingerprint generation unit (GFF).

Using the neural fingerprint lookup-table (FPLT) improves the adaptive matching unit (AMU) via combining the two fingerprints (coarse and fine), then it matches them with the stored signatures, or if it is a newly discovers fingerprint, it stores the fingerprint into the FPLT and a new entry. The building blocks of the Zydeco-Style architecture are:

\begin{itemize}
	\item \textit{Spike Fingerprint Unit:} is responsible for generating the fine fingerprint of the firing neuron. It consists of three subunits units: adaptive threshold spike detector, Pivot Finder, and Fingerprint Generator.
	\item \textit{Global Analysis Unit:} is responsible for the time of the dominant spike analysis, where the delay of the spike as seen from the surrounding channels is calculated and the spiking frequency analysis for neural identification. In order to generate the global-local fingerprints.
	\item \textit{Adaptive Matching Unit:} matches the current fingerprint value and stores the pre-loaded population of the fingerprints at the FPLT. It consists of an Artificial Immunity System matching unit (AIS) and 32Kb of SRAM memory, which is used to store the initial population of fingerprints as well as a working space for the matching unit.
\end{itemize}

\begin{figure}
	\centerline{\includegraphics[width=8.7cm, height= 5.5 cm]{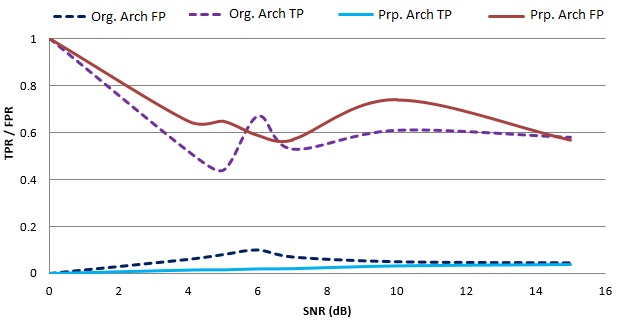}}
	\caption{The result comparison graph shows the proposed architecture performance (in solid lines) and previous architecture (in dotted lines).}
	\label{figure6}
\end{figure}

\section{Simulation Results}
For the model validation and verification, the proposed architecture was simulated and implemented in Matlab and Verilog, respectively. Where we used two groups of simulated data sets, three for the training phase (SNR of 0dB and 5dB) and another two for testing with a different SNR (7dB and 10dB); the dataset was created based on the method described in~\cite{12}. The initial population of the neural fingerprints was set via the ICA classification function in WaveClus$\copyright
$ based on the open-source algorithm for spike sorting available at~\cite{16} and~\cite{17}. As a result, the detection accuracy is optimal in the 0dB dataset case. On the other hand, the system's performance accuracy degrades at SNR $>$ 7dB cases. However, when we compare our results to the architecture described in Section II; using the same approach as in~\cite{5}, where we used a pair of {False-Positive Rates (FPR), True-Positive Rates (TPR)} as a system performance measurement parameter for neural detection, where the optimal detection must generate {0, 1} values. The results comparison is promising, as shown in Fig.~\ref{figure6}.

From the graph, we can see that the system improves the detection TPR rate by 23\% (due to the added global-local fingerprint unit) and reduces rejection FPR rates by 96.3\% (due to the negative selection algorithm). However, this comes at the cost of an additional fingerprint unit (34.5\% increase in chip area); on the other hand, we achieve a higher (104\%) detection accuracy rate. Moreover, due to the enhanced TPR performance, the detection unit can be used in a broader range of medical applications and still satisfy the implant limitations. Further, we are aware that performance can be improved even more via a proper selection of the initial fingerprint population.

\section{Conclusion and Future Work}
\begin{figure}
	\centerline{\includegraphics[width=8.5cm, height= 4cm]{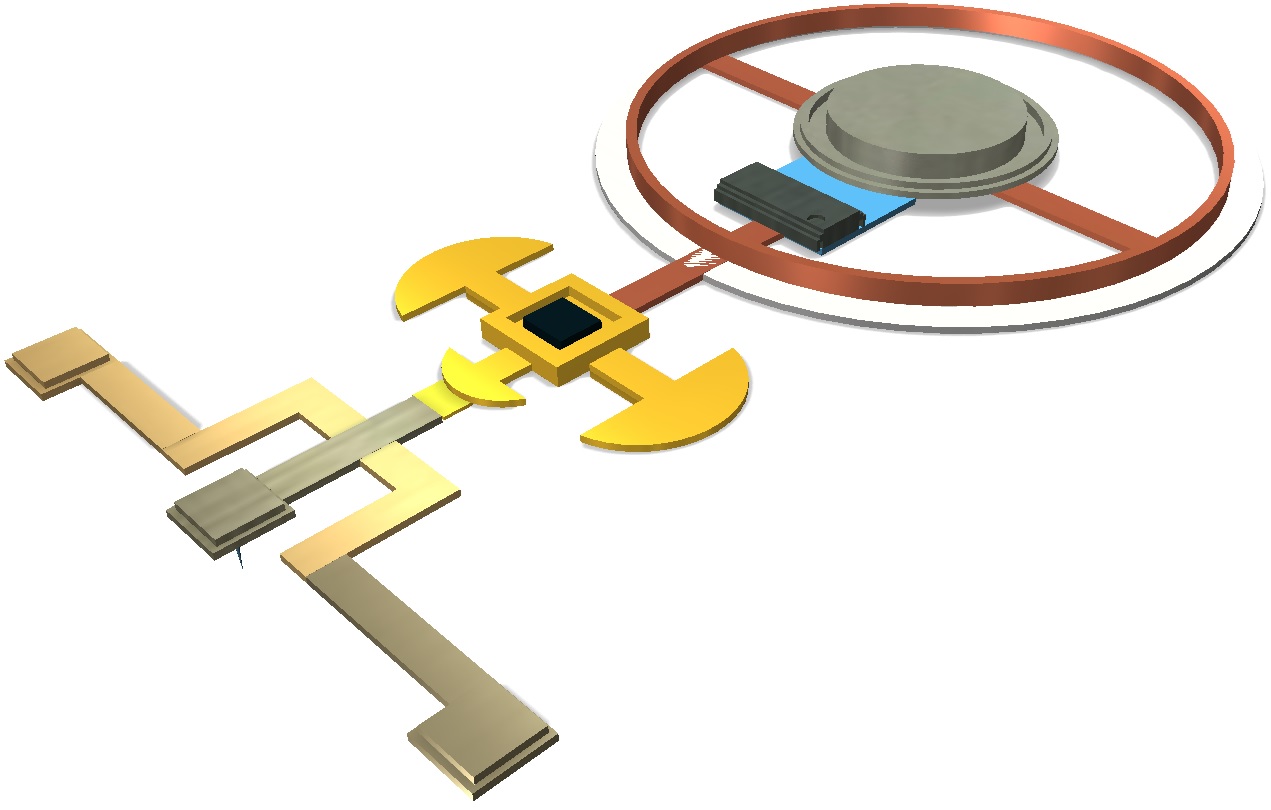}}
	\caption{The Z-Plant\copyright~Concept design diagram, the three components are (from left to right): Microelectrode array, AFE and Spike Sorter chip, and the I/O coil with the  RF communication unit and the SRAM chip diagram with the three components from left to right: electroarray, AFE and Spike Sorter chip, and the I/O RF communication unit and the SRAM chip.}
	\label{figure7}
\end{figure}

The proposed neural connectivity model blended with the Zydeco-Style spike sorting architecture gives a new way of detection perspective for analyzing the firing neural fingerprints as well as a novel neural analysis vision of the surrounding signaling activity of the adjacent neurons in time and frequency domains. The proposed architecture gives a better detection result and opens new opportunities for medical IoT to continue monitoring mental disorders and early detection tasks. The proposed model can contribute to a new era of adaptive real-time BCI implants, especially when coupled with the proposed layered implant, which consists of three parts: I/O and battery unit, the spike sorter and AFE, and the electrode array. Each of these units is implanted in a different layer of the skull: under the skin, over the dura, and inside the Pia matter itself. The concept implant design is shown in Fig.~\ref{figure7}. Finally, for future work, we will implement the Zydeco-Style VLSI spike sorting in ASIC architecture and study the required enhancements to enable neural activity prediction features for mental disorders.

\bibliographystyle{ieeetr}
\bibliography{zydacoReferences}
\end{document}